\documentclass[10pt]{article}

\usepackage[T2A]{fontenc}
\usepackage{latexsym}
\usepackage{amssymb}
\usepackage{amsmath}

\def\R{{\Bbb R}}

\def\T{{\rm T}}

\def\Mat{\mathrm{Mat}}

\def\diag{{\rm diag}}

\newcommand{\be}{\begin{equation}}
\newcommand{\ee}{\end{equation}}

\def\U{{\rm U}}

\begin{document}

\title{Gauge invariant generalizations of the Proca equation and the Yang-Mills-Proca equation}

\maketitle


\author{Nikolay Marchuk\footnote{1) Steklov Mathematical Institute of Russian Academy of Sciences, Moscow

2) HSE University, Moscow

orcid.org/0000-0001-6185-434X}
}

\begin{abstract}
The Proca equations (1936) are used in quantum field theory to describe vector bosons (spin 1) with a nonzero mass. The Proca equations are not gauge invariant. In contrast to Stueckelberg's approach (1938), this article presents a gauge invariant generalization of the Proca equation by introducing an additional vector field into the Proca equation. The results are extended to the Yang-Mills-Proca equations, leading to equations with non-Abelian gauge symmetry.
\end{abstract}


\medskip

MSC 70S15



\bigskip

The Proca equation \cite{Proca1936} was proposed in 1936 by the Romanian physicist Alexandru Proca as a generalization of Maxwell's equations to describe particles with spin 1 and a nonzero mass $m$. Unlike Maxwell's equations, which are invariant under the abelian gauge symmetry with  unitary group $\U(1)$, the Proca equations are not gauge-invariant.

In 1938, Stueckelberg \cite{Sh1938a,Sh1938b,Sh1938c} considered a gauge-invariant modification of the Proca equation obtained by introducing an additional scalar field into the Proca equation.

Later, a number of modifications and generalizations of the Proca equation were proposed (see \cite{BRST1}, \cite{BRST2}, \cite{BRST3}, \cite{gen:Proca}, \cite{gen:Proca:2} and a review \cite{Ruegg2004}).

This article considers a gauge-invariant generalization of the Proca equation, obtained by introducing an additional vector field into the Proca equation. The results are extended to the Yang-Mills-Proca equations and lead to equations with non-Abelian gauge symmetry. Also we consider a case of several masses. 

We hope that the proposed formalism has the potential to become an improved analogue of the standard model of electroweak interaction within the framework of the project being developed by the author \cite{Mar1},
 \cite{Mar2}, \cite{Mar3}, \cite{Mar4}.

All physical constants (speed of light, Planck's constant, positron charge, etc.), except for particle masses, are taken to be equal to one.


\section{Maxwell's equations (1862) and Proca's equations (1936)}

In this work, all equations are considered in Minkowski space $\R^{1,3}$ with Cartesian coordinates $x^\mu$, $\mu=0,1,2,3$, with partial derivatives $\partial_\mu=\partial/\partial x^\mu$, and with a metric tensor given by the diagonal matrix $\eta=\|\eta^{\mu\nu}\|=\|\eta_{\mu\nu}\|=\diag(1,-1,-1,-1)$ of the Lorentz signature. In the components of tensor fields, indices are raised/lowered using the metric tensor. The Einstein summation convention is used for repeated indices.

\medskip

\noindent{\bf Maxwell's equations in relativistic form.} Real tensor fields in Minkowski space are considered. Let $a_\mu$ be the co-vector potential of the electromagnetic field, $f_{\mu\nu}$ be the skew-symmetric covariant tensor of the electromagnetic field strength, and $j^\nu$ be the electric current vector. Maxwell's equations in relativistic form are written as \cite{Novikov:Taymanov}
\begin{eqnarray}
\partial_\mu a_\nu - \partial_\nu a_\mu &=& f_{\mu\nu},\quad \mu,\nu=0,1,2,3,\label{Maxw1}\\
\partial_\mu f^{\mu\nu} &=& j^\nu,\quad \nu=0,1,2,3.\label{Maxw2}
\end{eqnarray}
Substituting $f_{\mu\nu}$ from the first equation into the second, we obtain
\begin{equation}
\square a^\nu - \partial^\nu(\partial_\mu a^\mu)=j^\nu,\label{sqaj}
\end{equation}
where $\square=\partial_\mu\partial^\mu$ is the d'Alembert operator.

A consequence of equation (\ref{sqaj}) is the equality (the 4-divergence of the current vector is equal to zero)
\begin{equation}
\partial_\nu j^\nu=0.\label{divj0}
\end{equation}

The Maxwell system of equations (\ref{Maxw1}),(\ref{Maxw2}) is invariant under the gauge transformation (symmetry)
\begin{eqnarray}
&& a_\mu\to\acute a_\mu=a_\mu-\partial_\mu\lambda,\label{Gauge:afj:1}\\
&& f_{\mu\nu}\to\acute f_{\mu\nu}=f_{\mu\nu},\label{Gauge:afj:2}\\
&& j^\nu\to\acute j^\nu=j^\nu,\label{Gauge:afj:3}
\end{eqnarray}
where $\lambda=\lambda(x)$ is a smooth\footnote{In this article, we assume that the smoothness of all functions (tensor fields) is sufficient for the validity of our arguments.} real valued function $\R^{1,3}\to\R$. The presence of gauge symmetry allows us to impose certain conditions (calibration) on the fields entering the equations (\ref{Maxw1}),(\ref{Maxw2}) without changing the physical meaning of the equations.  In particular, the condition is often used
\begin{equation}
\partial_\mu a^\mu=0,\label{lor:gauge}
\end{equation}
which is called the {\em Lorentz gauge}. Adding the Lorentz gauge to the equations (\ref{Maxw1}),(\ref{Maxw2}) gives
\begin{equation}
\square a^\nu = j^\nu, \quad \nu=0,1,2,3.\label{dal:a}
\end{equation}
\medskip

\noindent{\bf Proca's equation (1936).} The Proca equation \cite{Proca1936} is considered as the following modification of Maxwell's equations:
\begin{eqnarray}
\partial_\mu a_\nu - \partial_\nu a_\mu &=& f_{\mu\nu},\quad \mu,\nu=0,1,2,3,\label{Maxw1:P}\\
\partial_\mu f^{\mu\nu} + m^2 a^\nu &=& 0,\quad \nu=0,1,2,3,\label{Maxw2:P}
\end{eqnarray}
where $m$ is a real nonzero constant, which is interpreted as the mass of the vector boson. A consequence of equations (\ref{Maxw1:P}), (\ref{Maxw2:P}) will be the equality
$$
m^2\partial_\nu a^\nu=0,
$$
which, when $m\neq 0$, gives the Lorentz gauge (\ref{lor:gauge}) and the Klein-Gordon equation for the vector potential $a^\nu$
\begin{equation}
(\square+m^2)a^\nu = 0, \quad \nu=0,1,2,3.\label{dal:a:m}
\end{equation}

The Proca equation (\ref{Maxw1:P}),(\ref{Maxw2:P}) does not have a gauge symmetry.
\medskip

\noindent{\bf Inhomogeneous Proca equation.} Let us add a vector $j^\nu$ to the right-hand side of equations (\ref{Maxw1:P}), (\ref{Maxw2:P}). We obtain the inhomogeneous equations
\begin{eqnarray}
\partial_\mu a_\nu - \partial_\nu a_\mu &=& f_{\mu\nu},\quad \mu,\nu=0,1,2,3,\label{Maxw1:PN}\\
\partial_\mu f^{\mu\nu} + m^2 a^\nu &=& j^\nu,\quad \nu=0,1,2,3,\label{Maxw2:PN}
\end{eqnarray}
from which it follows the condition
\begin{equation}
\partial_\nu j^\nu - m^2\partial_\nu a^\nu=0\label{Jma}
\end{equation}
and the inhomogeneous Klein-Gordon equation
\begin{equation}
(\square+m^2)a^\nu = j^\nu + \frac{1}{m^2}\partial^\nu(\partial_\mu j^\mu).\label{KG:j}
\end{equation}


\section{A gauge invariant generalized Proca equation}

The equations (\ref{Maxw1:PN}),(\ref{Maxw2:PN}) do not have gauge symmetry (\ref{Gauge:afj:1}), (\ref{Gauge:afj:2}), (\ref{Gauge:afj:3}), but they have the following gauge symmetry:
\begin{eqnarray}
&& a_\mu\to\acute a_\mu=a_\mu-\partial_\mu\lambda,\label{Gauge:afj:1x}\\
&& f_{\mu\nu}\to\acute f_{\mu\nu}=f_{\mu\nu},\label{Gauge:afj:2x}\\
&& j^\nu\to\acute j^\nu=j^\nu - m^2\partial^\nu\lambda.\label{Gauge:afj:3x}
\end{eqnarray}
If we introduce the new notation
$$
k^\nu := \frac{1}{m^2}j^\nu,
$$
then, by (\ref{Gauge:afj:3x}), the vector $k^\nu$ transforms according to the rule
$$
k^\nu\to\acute k^\nu= k^\nu-\partial^\nu\lambda.
$$
In geometry, such a transformation is usually associated with a connection, and in field theory, with a vector potential.
Where might the vector potential $k^\nu$ come from?

Our proposal is to take the vector $k^\nu$ to satisfy the homogeneous Maxwell equations
$$
\partial_\mu k_\nu-\partial_\nu k_\mu = g_{\mu\nu},\quad \partial_\mu g^{\mu\nu}=0.
$$
So, we arrive at the system of equations
\begin{eqnarray}
\partial_\mu a_\nu - \partial_\nu a_\mu &=& f_{\mu\nu},\label{Maxw1:PNx}\\
\partial_\mu f^{\mu\nu} + m^2 a^\nu &=& m^2 k^\nu,\label{Maxw2:PNx}\\
\partial_\mu k_\nu-\partial_\nu k_\mu &=& g_{\mu\nu},\label{Maxw3:PNx}\\
\partial_\mu g^{\mu\nu} &=& 0,\label{Maxw4:PNx}
\end{eqnarray}
which is invariant with respect to the gauge transformation
\begin{eqnarray}
&& a_\mu\to a_\mu-\partial_\mu\lambda,\quad k^\nu\to k^\nu - \partial^\nu\lambda.\label{Gauge:afj:1xx}\\
&& f_{\mu\nu}\to f_{\mu\nu},\quad g_{\mu\nu}\to g_{\mu\nu},\label{Gauge:afj:2xx}
\end{eqnarray}
where $\lambda : \R^{1,3}\to \R$. From the equations (\ref{Maxw1:PNx}), (\ref{Maxw2:PNx}) for $m\neq 0$ we obtain the following consequence
$$
\partial_\nu a^\nu = \partial_\mu k^\mu.
$$
Therefore, the Lorentz gauge $\partial_\nu a^\nu =0$ for the vector potential $a^\nu$ implies the Lorentz gauge $\partial_\nu k^\nu =0$ for the vector potential $k^\nu$. And, in the case of the Lorentz gauge, we have
\begin{equation}
(\square+ m^2)a^\nu= m^2 k^\nu,\quad \square k^\nu=0.\label{KG:ak}
\end{equation}

For equations (\ref{KG:ak}) we distinguish two classes of solutions. First,
\begin{equation}
(\square+ m^2)a^\nu=0,\quad k^\nu=0.\label{sqm0}
\end{equation}
Secondly,
\begin{equation}
\square a^\nu=0,\quad \square k^\nu=0,\quad a^\nu=k^\nu.\label{sqm1}
\end{equation}
We are interested in the solutions (\ref{sqm0}) since they depend on the mass $m$ and are solutions of the Proca equation
(\ref{Maxw1:P}), (\ref{Maxw2:P}).
\medskip

\noindent{\bf Conclusion.} The gauge-invariant system of equations (\ref{Maxw1:PNx}), (\ref{Maxw2:PNx}), (\ref{Maxw3:PNx}), (\ref{Maxw4:PNx}) contains among its solutions {\em all} solutions of the Proca equation (\ref{Maxw1:P}), (\ref{Maxw2:P}). And, therefore, can be considered as a gauge-invariant {\em generalization of the Proca equation}.
\medskip

\noindent{\bf Remark.} It is easy to see that, due to the linearity of Maxwell's and Proca's equations, the system of equations (\ref{Maxw1:PNx})-(\ref{Maxw4:PNx}) splits into two independent systems of equations. Indeed, from equations (\ref{Maxw2:PNx}), (\ref{Maxw4:PNx}), we obtain the equation
$$
\partial_\mu(f^{\mu\nu}-g^{\mu\nu}) + m^2(a^\nu-k^\mu)=0.
$$
From equations (\ref{Maxw1:PNx}), (\ref{Maxw3:PNx}), we obtain
$$
\partial_\mu(a_\nu-k_\nu)-\partial_\nu(a_\mu-k_\mu)=f_{\mu\nu}-g_{\mu\nu}.
$$
Therefore, the system of equations (\ref{Maxw1:PNx})-(\ref{Maxw4:PNx}) can be rewritten as a system consisting of two independent systems of equations (for $\tilde a_\mu=a_\mu-k_\mu$, $\tilde f_{\mu\nu}=f_{\mu\nu}-g_{\mu\nu}$, $k_\mu$, $g_{\mu\nu}$)
\begin{eqnarray}
\partial_\mu \tilde a_\nu-\partial_\nu \tilde a_\mu=\tilde f_{\mu\nu},\quad
\partial_\mu \tilde f^{\mu\nu}+m^2 \tilde a^\nu=0,\label{eqt:1}\\
\partial_\mu k_\nu-\partial_\nu k_\mu=g_{\mu\nu},\quad
\partial_\mu g^{\mu\nu}=0\label{eqt:2}
\end{eqnarray}
with the gauge symmetry
$$
\tilde a_\mu\to\tilde a_\mu,\quad \tilde f_{\mu\nu}\to\tilde f_{\mu\nu},\quad
k_\mu\to k_\mu-\partial_\mu\lambda,\quad g_{\mu\nu}\to g_{\mu\nu}.
$$
For the nonlinear generalized Yang-Mills-Proca equations considered below, the statement that the system of equations splits into two independent subsystems of equations is not true. The systems of equations with several masses considered below, generally speaking, also do not split into independent subsystems.


\section{Generalization of Proca's equation to the case of several masses}

Let us consider $n\geq 2$ real vector potentials
$$
a_\mu(k),\quad k=1,\ldots,n
$$
satisfying a system of equations depending on $n-1$ real numbers (masses) $m_1,\ldots,m_{n-1}\in\R$
\begin{eqnarray}
\partial_\mu a_\nu(k) - \partial_\nu a_\mu(k) &=& f_{\mu\nu}(k),\label{Maxw0:PNxy}\\
\partial_\mu f^{\mu\nu}(k) + m_k{}^2 a^\nu(k) &=& m_k{}^2 a^\nu(k+1),\quad k=1,\ldots,n-1\label{Maxw1:PNxy}\\
\partial_\mu a_\nu(n) - \partial_\nu a_\mu(n) &=& f_{\mu\nu}(n),\quad
\partial_\mu f^{\mu\nu}(n) = j^\nu\label{Maxw2:PNxy}
\end{eqnarray}
with the gauge symmetry
\begin{eqnarray}
&& a_\mu(k)\to a_\mu(k)-\partial_\mu\lambda,\quad k=1,\ldots,n,\label{gau:afj1}\\
&& f_{\mu\nu}(k)\to f_{\mu\nu}(k),\\
&& j^\nu\to j^\nu.
\end{eqnarray}
For $m_1,\ldots,m_{n-1}\neq 0$ the system of equations (\ref{Maxw0:PNxy}), (\ref{Maxw1:PNxy}), (\ref{Maxw2:PNxy}) has the consequences
\begin{equation}
\partial_\nu a^\nu(1) = \cdots = \partial_\nu a^\nu(n),\quad \partial_\nu j^\nu=0.\label{aaa:j}
\end{equation}
Consider the Lorentz gauge $\partial_\nu a^\nu(1)=0$ for the vector $a^\nu(1)$. By (\ref{aaa:j}), we obtain
\begin{equation}
\partial_\nu a^\nu(1)=\cdots=\partial_\nu a^\nu(n)=0\label{aaaa}
\end{equation}
and equations (\ref{Maxw0:PNxy}), (\ref{Maxw1:PNxy}), (\ref{Maxw2:PNxy}) are written as
\begin{eqnarray}
(\square+ m_k{}^2)a^\nu(k) &=& m_k{}^2 a^\nu(k+1),\quad k=1,\ldots,n-1,\label{ssqq}\\
\square a^\nu(n) &=& j^\nu\nonumber
\end{eqnarray}

\medskip

To take the next step in generalizing the equations (\ref{Maxw0:PNxy}), (\ref{Maxw1:PNxy}), (\ref{Maxw2:PNxy}) we introduce the Euclidean space $\R^n$.
We will write its elements as $n$-dimensional columns. We denote by $\hat e_1,\ldots,\hat e_n$ the vectors of the orthonormal basis of the space $\R^n$ ($\hat e_k$ is the $n$-dimensional column with a one at the $k$-th place and zeros at the remaining places).

We will use tensors (tensor fields) of the Minkowski space $\R^{1,3}$
(we use small Greek letters to denote the tensor indices of the Minkowski space  $\mu,\nu,\ldots = 0,1,2,3$) with values in  the Euclidean space $\R^n$
$$
\hat a_\mu = \left[\begin{array}{c} a_\mu(1)\\ \vdots\\ a_\mu(n)\end{array}\right],\quad
\hat f_{\mu\nu} = \left[\begin{array}{c} f_{\mu\nu}(1)\\ \vdots\\ f_{\mu\nu}(n)\end{array}\right],\quad
\hat j^\nu = \left[\begin{array}{c} j^\nu(1)\\ \vdots\\ j^\nu(n)\end{array}\right],
$$
$$
\hat a_\mu =\sum_{k=1}^n a_\mu(k)\hat e_k,\quad
\hat f_{\mu\nu} =\sum_{k=1}^n f_{\mu\nu}(k)\hat e_k,\quad
\hat j^\nu =\sum_{k=1}^n j^\nu(k)\hat e_k,
$$
Let us denote
$$
\hat 1 = \sum_{k=1}^n \hat e_k =
\left[\begin{array}{c} 1\\ \vdots\\ 1\end{array}\right],\quad
\hat 0 =
\left[\begin{array}{c} 0\\ \vdots\\ 0\end{array}\right].
$$
Let's introduce a matrix $M\in\Mat(n,\R)$ such that the sum of the elements of each of its rows is equal to zero. This condition can be written as follows:
\begin{equation}
M\hat 1=\hat 0.\label{M10}
\end{equation}
In other words, the matrix $M$ has the eigenvector $\hat 1$ corresponding to the zero eigenvalue ($\det M=0$).

The system of equations generalizing the system of equations (\ref{Maxw0:PNxy}), (\ref{Maxw1:PNxy}), (\ref{Maxw2:PNxy}) is written as
\begin{eqnarray}
&& \partial_\mu\hat a_\nu - \partial_\nu\hat a_\mu = \hat f_{\mu\nu},\label{GProca:M1}\\
&& \partial_\mu\hat f^{\mu\nu} + M\hat a^\nu = \hat j^\nu.\label{GProca:M2}
\end{eqnarray}
Condition (\ref{M10}) ensures gauge invariance of the system of equations (\ref{GProca:M1}), (\ref{GProca:M2}) with respect to the transformation ($\lambda=\lambda(x) : \R^{1,3}\to\R$)
$$
a_\mu(k) \to a_\mu(k)-\partial_\mu\lambda,\quad f_{\mu\nu}(k)\to f_{\mu\nu}(k),\quad j^\nu(k)\to j^\nu(k),\quad k=1,\ldots,n.
$$

We also note that for $n\to\infty$, instead of the Euclidean space $\R^n$, we can consider the Hilbert space ${\cal H}$, and the tensor fields of the Minkowski space $\hat a_\mu$, $\hat f_{\mu\nu}$, $\hat j^\nu$ included in the system of equations (\ref{GProca:M1}), (\ref{GProca:M2}) can be considered as tensor fields of the Minkowski space with values in the Hilbert space ${\cal H}$.


\section{Gauge invariant generalized Yang-Mills-Proca equations}

\noindent{\bf Yang-Mills equations.} Let $G$ be a unitary or semisimple Lie group and $L$ be its real Lie algebra. In the Minkowski space $\R^{1,3}$ with Cartesian coordinates $x^\mu$, $\mu=0,1,2,3$, we consider tensor fields with values in the Lie algebra $L$.
The Yang-Mills equations \cite{Yang:Mills1954}, \cite{Novikov:Taymanov} are a system of equations
\begin{eqnarray}
&& \partial_\mu A_\nu - \partial_\nu A_\mu - [A_\mu, A_\nu] = F_{\mu\nu},\label{YM1}\\
&& \partial_\mu F^{\mu\nu} - [A_\mu, F^{\mu\nu}] = J^\nu,\label{YM2}
\end{eqnarray}
where $A_\mu\in L\T_1$ is the Yang-Mills field potential, $F_{\mu\nu}=-F_{\nu\mu}\in L\T_2$ is the Yang-Mills field strength, $J^\nu\in L\T^1$ is the (non-Abelian) current vector, and the commutator of elements of the Lie algebra $L$ is denoted by square brackets.

It can be verified that from equations (\ref{YM1}), (\ref{YM2}) follows the corollary
\begin{equation}
\partial_\nu J^\nu - [A_\nu, J^\nu] =0.\label{con:YM}
\end{equation}

The system of equations (\ref{YM1}), (\ref{YM2}) is invariant under the gauge transformation
\begin{eqnarray}
A_\mu \to \acute A_\mu &=& U^{-1}A_\mu U - U^{-1}\partial_\mu U,\nonumber\\
F_{\mu\nu} \to \acute F_{\mu\nu} &=& U^{-1}F_{\mu\nu} U,\label{AFJ:trans}\\
J^{\mu} \to \acute J^{\mu} &=& U^{-1}J^{\mu} U,\nonumber
\end{eqnarray}
where $U : \R^{1,3}\to G$ is an arbitrary smooth function with values in the Lie group $G$.
\medskip

\noindent{\bf The Yang-Mills-Proca equations} have the form \cite{MarShir2016}, \cite{Shir2022}
\begin{eqnarray}
&& \partial_\mu A_\nu - \partial_\nu A_\mu - [A_\mu, A_\nu] = F_{\mu\nu},\label{YMP1}\\
&& \partial_\mu F^{\mu\nu} - [A_\mu, F^{\mu\nu}] + m^2 A^\nu = 0,\label{YMP2}
\end{eqnarray}
where $A_\mu\in L\T_1$, $F_{\mu\nu}=-F_{\nu\mu}\in L\T_2$, $m\in\R$, $m\neq 0$.

The system of equations (\ref{YMP1}), (\ref{YMP2}) is not gauge invariant under the transformation (\ref{AFJ:trans}).
\medskip

\noindent{\bf Generalized Yang-Mills-Proca equations}. Let $n\geq 2$ be an integer number. We consider sets of $n$ tensor fields of Minkowski space
 $A_\mu(1),\ldots,A_\mu(n)\in L\T_1$; $F_{\mu\nu}(1),\ldots,F_{\mu\nu}(n)\in L\T_2$; $J^\mu(1),\ldots,J^\mu(n)\in L\T^1$
 with values in the Lie algebra
 $L$. The generalized Yang-Mills-Proca equations are written as
\begin{eqnarray}
&& \partial_\mu \hat A_\nu - \partial_\nu \hat A_\mu - [\hat A_\mu, \hat A_\nu] = \hat F_{\mu\nu},\label{YMP1z}\\
&& \partial_\mu \hat F^{\mu\nu} - [\hat A_\mu, \hat F^{\mu\nu}] + M \hat A^\nu = \hat J^\nu,\label{YMP2z}
\end{eqnarray}
where
$$
\hat A_\mu = \left[\begin{array}{c} A_\mu(1)\\ \vdots\\ A_\mu(n)\end{array}\right],\quad
\hat F_{\mu\nu} = \left[\begin{array}{c} F_{\mu\nu}(1)\\ \vdots\\ F_{\mu\nu}(n)\end{array}\right],\quad
\hat J^\nu = \left[\begin{array}{c} J^\nu(1)\\ \vdots\\ J^\nu(n)\end{array}\right],
$$
and the matrix $M\in\Mat(n,\R)$ is such that the sum of the elements of each of its rows is equal to zero.

The system of equations (\ref{YMP1z}), (\ref{YMP2z}) is invariant under the gauge transformation ($k=1,\ldots,n$)
\begin{eqnarray}
A_\mu(k) \to \acute A_\mu(k) &=& U^{-1}A_\mu(k) U - U^{-1}\partial_\mu U,\nonumber\\
F_{\mu\nu}(k) \to \acute F_{\mu\nu}(k) &=& U^{-1}F_{\mu\nu}(k) U,\label{AFJ:trans1}\\
J^{\mu}(k) \to \acute J^{\mu}(k) &=& U^{-1}J^{\mu}(k) U,\nonumber
\end{eqnarray}
where $U : \R^{1,3}\to G$ is an arbitrary smooth function with values in the Lie group $G$.
\medskip

\noindent{\bf Example for $n=2$.} Let us take the matrix $M$ in the form
$$
M = \left[\begin{array}{cc} m_1{}^2 & -m_1{}^2 \\ -m_2{}^2 & m_2{}^2\end{array}\right],
$$
where $m_1,m_2\in\R$. In this case, the generalized system of Yang-Mills-Proca equations (\ref{YMP1z}), (\ref{YMP2z}) for $\hat J^\nu=0$ can be written as
\begin{eqnarray}
&& \partial_\mu A_\nu(1) - \partial_\nu A_\mu(1) - [A_\mu(1), A_\nu(1)] = F_{\mu\nu}(1),\label{YMP1zt1}\\
&& \partial_\mu A_\nu(2) - \partial_\nu A_\mu(2) - [A_\mu(2), A_\nu(2)] = F_{\mu\nu}(2),\label{YMP1zt2}\\
&& \partial_\mu F^{\mu\nu}(1) - [A_\mu(1), F^{\mu\nu}(1)] + m_1{}^2 A^\nu(1) - m_1{}^2 A^\nu(2) = 0,\label{YMP2zt1}\\
&& \partial_\mu F^{\mu\nu}(2) - [A_\mu(2), F^{\mu\nu}(2)] - m_2{}^2 A^\nu(1) + m_2{}^2 A^\nu(2) = 0.\label{YMP2zt2}
\end{eqnarray}
By transferring the terms with masses to the right-hand side of the equations (\ref{YMP2zt1}), (\ref{YMP2zt2}), for $m_1\neq 0$, $m_2\neq 0$ we obtain the consequences of the equations
\begin{eqnarray}
&& \partial_\nu(-A^\nu(1)+A^\nu(2)) - [A_\nu(1), -A^\nu(1)+A^\nu(2)] = 0,\label{cons:10}\\
&& \partial_\nu(A^\nu(1)-A^\nu(2)) - [A_\nu(2), A^\nu(1)-A^\nu(2)] = 0.\label{cons:20}
\end{eqnarray}
It is easy to see that corollaries (\ref{cons:10}) and (\ref{cons:20}) are equivalent.

Equations (\ref{YMP1zt1}), (\ref{YMP1zt2}), (\ref{YMP2zt1}), (\ref{YMP2zt2}) can be written as equations for potentials $A_\mu(1)$, $A_\mu(2)$
\begin{align*}
\partial_\mu( \partial^\mu A^\nu(1) - \partial^\nu A^\mu(1)
 - [A^\mu(1), A^\nu(1)])  &-\\
  [A_\mu(1), \partial^\mu A^\nu(1) - \partial^\nu A^\mu(1) - [A^\mu(1), A^\nu(1)]] &+
 m_1{}^2 A^\nu(1) = m_1{}^2 A^\nu(2),\\
 \partial_\mu( \partial^\mu A^\nu(2) - \partial^\nu A^\mu(2) - [A^\mu(2), A^\nu(2)]) &-\\
 [A_\mu(2), \partial^\mu A^\nu(2) - \partial^\nu A^\mu(2) - [A^\mu(2), A^\nu(2)]] &+
 m_2{}^2 A^\nu(2) = m_2{}^2 A^\nu(1).
 \end{align*}
\medskip

\noindent{\bf Conclusion.} Thus, we have considered a gauge-invariant generalization of the Proca equation (\ref{GProca:M1}), (\ref{GProca:M2}) with an Abelian gauge symmetry. We have also considered a gauge-invariant generalization of the Yang-Mills-Proca equations (\ref{YMP1z}), (\ref{YMP2z}) with a non-Abelian gauge symmetry.
\medskip

\noindent{\bf Acknowledgments.}
This work is supported by the Russian Science Foundation (project 23-71-10028-P), https://rscf.ru/en/project/23-71-10028-P/.

 \end{document}